\documentclass[10pt,conf,compsoc]{IEEEtran}
\usepackage{hyperref}
\usepackage[numbers]{natbib}

\usepackage{microtype}
\usepackage{amsthm}
\usepackage{graphicx}
 \usepackage[font={small}]{caption, subfig}
\usepackage{balance}
\usepackage{epstopdf}
\usepackage{mathtools}
\usepackage{hyperref}
\usepackage{nameref}
\usepackage{enumerate}
\usepackage{paralist}
\usepackage{tabularx}
\usepackage{float}
\usepackage{xcolor, colortbl}
\usepackage{multirow}
\usepackage{xintexpr}
\usepackage{sober}

\usepackage[ruled,vlined,algo2e]{algorithm2e} 

\newcommand{\figwidth}{.9}
\newcommand{\subfigwidth}{.45}
\newcommand{\threesubfigwidth}{.32}

\newcommand{\bannerwidth}{.7}
\newcommand{\boxscale}{.9}

\newcommand{\topk}{10}
\newcommand{\twotopk}{{\xinttheexpr 2*\topk{}\relax}}
\DeclareMathOperator*{\iqr}{iqr}
\DeclareMathOperator*{\median}{med}

\DeclareMathOperator*{\cost}{bytes}
\DeclareMathOperator*{\correctbase}{measure}
\DeclareMathOperator*{\correct}{correct}

\DeclareMathOperator*{\argmax}{argmax}
\SetKwInput{KwGiven}{Given}
\SetKwInput{KwFind}{Select}
\SetKwInput{KwWhere}{Where}
\SetKwInput{KwMin}{Minimizing}
\newcommand{\bfs}{\texttt{bfs}}
\newcommand{\cluster}{\texttt{cluster}}
\newcommand{\degree}{\texttt{degree-net}}
\newcommand{\activity}{\texttt{activity-net}}
\newcommand{\random}{\texttt{random}}
\newcommand{\knns}{$\texttt{KNN}_s$}

\newcommand{\networkinference}{\mathcal{M}}

\newcommand{\networkquery}{\mathcal{W}}
\newcommand{\networkqueryr}{\mathcal{W}_\networkinference}
\newcommand{\classifier}{\mathcal{C}}
\newcommand{\classifierr}{\mathcal{C}_r}
\newcommand{\classifierset}{C_r}
\newcommand{\degreeflat}{\texttt{degree-top}}
\newcommand{\activityflat}{\texttt{activity-top}}
\newcommand{\ba}{BeerAdvocate}
\newcommand{\lfm}{Last.fm}
\newcommand{\ml}{MovieLens}
\newcommand{\knn}{\texttt{KNN}}
\newcommand{\tth}{\texttt{TH}}
\newcommand{\social}{\texttt{social}}
\newcommand{\bigo}{\mathcal{O}}
\newcommand{\netcost}{\cost(\networkqueryr)}
\newcommand{\taskcost}{\cost(\classifierr)}
\newcommand{\taskcorrect}{\correct(\classifierr)}
\newcommand{\efficiency}{\mathcal{E}(\networkqueryr)}

\SetAlFnt{\small}
\SetAlCapFnt{\small}
\SetAlCapNameFnt{\small}

\definecolor{LightCyan}{rgb}{0.88,1,1}
\definecolor{Gray}{gray}{0.8}
\definecolor{moss}{HTML}{006633}
\definecolor{midnight}{HTML}{000099}
\theoremstyle{definition}
\newtheorem{definition}{Definition}

\ifCLASSINFOpdf
\else
\fi
\begin{document}
%
\title{Inferring Network Structure From Data}
%
%
%
%

\author{Ivan Brugere,
        Tanya Berger-Wolf
\IEEEcompsocitemizethanks{\IEEEcompsocthanksitem Ivan Brugere is with the Department of Computer Science, University of Illinois at Chicago, E-mail: ibruge2@uic.edu

\IEEEcompsocthanksitem Tanya Berger-Wolf is at the Ohio State University and University of Illinois at Chicago, Email: berger-wolf.1@osu.edu}%
}

%
%

\markboth{}{}
%



\IEEEtitleabstractindextext{%
\begin{abstract}
Networks are complex models for underlying data in many application domains. In most instances, raw data is not natively in the form of a network, but derived from sensors, logs, images, or other data. Yet, the impact of the various choices in translating this data to a network have been largely unexamined. In this work, we propose a network model selection methodology that focuses on evaluating a network's utility for varying tasks, together with an efficiency measure which selects the most parsimonious model. We demonstrate that this network definition \textit{matters} in several ways for modeling the behavior of the underlying system. 
\end{abstract}


}
\maketitle

\IEEEdisplaynontitleabstractindextext

%
\IEEEpeerreviewmaketitle

\IEEEraisesectionheading{\section{Introduction}\label{sec:introduction}}

Networks are complex models for underlying data in many application domains. Networks model relationships between people, animals, genes, proteins, documents, media, language, products, etc. Often, we think of ``the network'' for an underlying system, for example, consider the social network of Zachary's karate club \cite{10.2307/3629752}. An edge in this network is defined as whether two individuals interacted outside of the club. But we cannot observe raw interaction data in terms of frequency, strength, missing data; many other networks from this data could have represented this classical network and a different network definition may yield different analyses and results on this network. In fact, most networks are not explicitly and unambiguously defined. For example, biological networks often measure correlations between the biological processes of cells, genes, or proteins. Human contact and social networks are often inferred from repeated interactions. These particular network definitions may also be very time-sensitive. Yet, they are often preprocessed, published, and analyzed as aggregated networks over time, which may produce different conclusions based on this choice of time scale \cite{Caceres2013,Holme201297}.

Multiple network repositories collect such preprocessed networks, largely without retaining the underlying raw data \cite{nr-aaai15, leskovec2016snap}. A network is not in and of itself a valuable representation for gaining a deeper understanding of the underlying system. Are these and other networks \textit{useful} representations for measuring and studying the behavior of the systems they model?

\begin{figure}
\centering
\includegraphics[width=\figwidth\columnwidth]{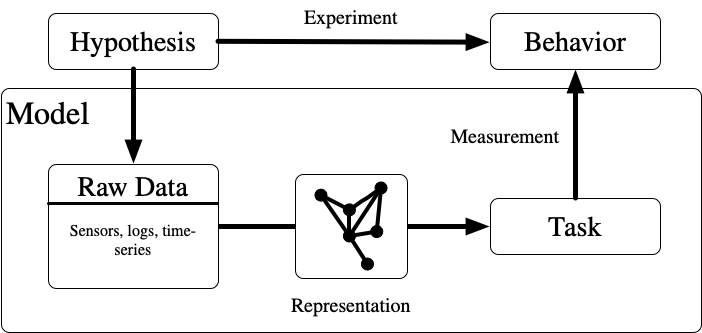}
\caption{A schematic overview of an observational science process, and its data-science modeling. Data-driven science models create representations (e.g. networks) on raw input data where edges or nodes are not natively and unambiguously defined. Much data science work focuses on the latter arc of this modeling to develop novel machine learning models to measure behavior on a given network structure.}
\label{fig:preschematic}
\end{figure}

Figure \ref{fig:preschematic} presents the traditional experimental science process and its corresponding data science modeling. Traditionally, experimentation is used to evaluate hypotheses and measure the behavior of interest in the underlying system. A data science model for this process constructs a representation over raw input data. This data is not natively in the form of a network and is subject to sampling/collection biases and limitations in measurement. These limitations, and other choices in network representation \textit{matter}. Prior work has largely ignored the challenges of translating raw data into a network representation and the downstream implications.\cite{brugere2018network} Instead, a great deal of work focuses on the latter piece of this modeling process: developing novel machine learning models on a \textit{given} network structure to measure behavior in the underlying system.

Why ought we construct a network representation in the first place? First, a network for a particular predictive task should perform better than an unstructured model which doesn't account for the relationships among entities (e.g. users). Evaluating network models against population methods measures whether there is a network effect at all within the underlying system. Also, high-order network properties (e.g. degree distribution, centrality, diffusion) are well studied. These properties can help generate hypotheses, and yield comparative analysis across diverse domains.

In this work, we propose a task-based network model selection methodology that compares multiple representations for predicting a particular behavior of the underlying system. We have two primary findings. First, we demonstrate that the best network definition is \textit{task-dependent}. That is, different network models are better representations for different tasks and there may not be one network representation for \textit{all} tasks. Second, in many instances, what we think of as network tasks are more \textit{efficiently} modeled by non-network representations such as clustering or population sampling.

Our strategy is to select or tune a network definition from a library of functions that transform the underlying data to a network representation. Work in network model selection has typically focused on inferring parameters of generative models from a given network, according to some structural or spectral features (e.g. degree distribution, eigenvalues) \cite{10.1371/journal.pone.0049949}. In contrast, we focus on selecting a representation that best performs a task or set of tasks. This current work combines and simplifies our prior methodological works for translating raw data to a network representation to perform particular tasks or behaviors \cite{brugereicdm2017, 10.1145/3184558.3191525}. 

We further extend this task-focused network model selection methodology, using minimum description length (MDL) criteria for selection. Our methodology measures predictive \textit{efficiency}, a \textit{general} and comparable measure of the network's performance of a local (i.e. node-level) predictive task of interest. This accounts for the complexity of both the network structure and the task predictor. Selection on efficiency favors parsimonious models and can be applied across arbitrary tasks and representations. We show \textit{stability, sensitivity, and significance testing} in our methodology.

\begin{figure*}[t]
\centering
  \includegraphics[width=\bannerwidth\textwidth]{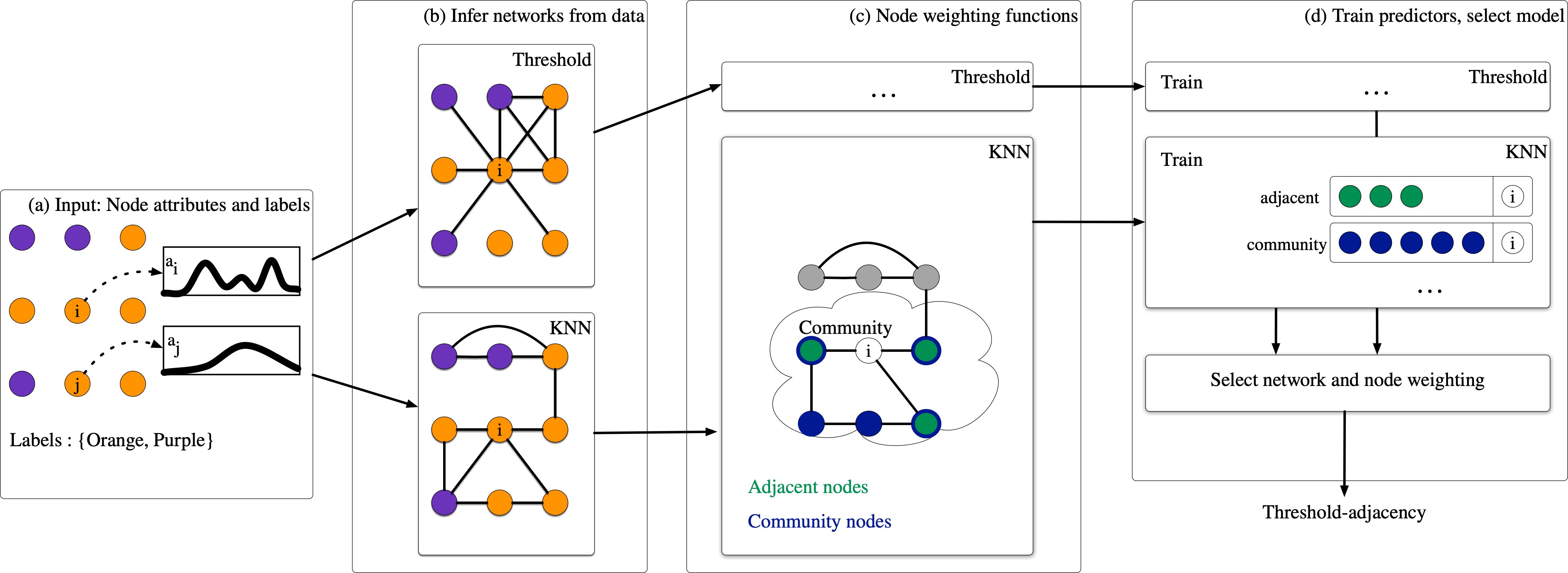}
  \caption{A high-level overview of our methodology. {\bf (a)} Given as input: nodes with associated attribute and label data. For example, attributes are some arbitrary distribution, and labels are a categorical value (Purple, Orange). {\bf (b)} On the input data, we define networks according to a collection of network models and their hyper-parameters. {\bf (c)} We generate node subsets based on various node weighting functions. Illustrated is the node subset corresponding to network adjacency  ({\color{moss}green}) and nodes within the same community as node $i$ ({\color{midnight}blue}). {\bf (d)} We train predictors on the node subsets generated by our node weighting function, for all nodes. We evaluate this set of predictors and select the network and node weighting function which maximizes our selection criteria. In this particular example, our framework selects the Threshold network, using adjacency weighting.} 
  \label{fig:overview}
\end{figure*}

\subsection{Related work}

Our work is primarily related to work in (1) network structure inference in networks (2) minimum-description length (MDL) approaches for networks.

Statistical relational learning in networks \cite{getoor20075} uses correlations between network structure, attribute distributions, and label distributions to build network-constrained predictive models. Our work uses two fundamental relational learning tasks, link prediction \cite{Hasan2011, Liben-Nowell2007} and collective classification \cite{namata:tkdd15} to evaluate network models inferred from data.

Network structure inference constructs network representations from data collected from sensors, online user trajectories or other underlying data \cite{brugere2018network, Kolaczyk2009}. Previous work has focused on evaluating these inferred network representations as predictive models for relational learning tasks. A `good' network in this setting is one which performs the task well, under some measure of robustness, cost, or stability. In this work, we combine and extend our previous work \cite{mlg2017_13, brugereicdm2017} examined model selection under varying network models, single or multiple tasks, and varying task methods.

Several generative models exist to model correlations between attributes, labels, and network structure and an attributed network from the inferred parameters. These include the Attributed Graph Model (AGM) \cite{Pfeiffer:2014:AGM:2566486.2567993}, the Multiplicative Attribute Graph Model (MAG) \cite{Kim2012}, and the Exponential Random Graph Model (ERGM) \cite{Robins2007}. These models are additive to our work and could be evaluated as candidate network inference models within our methodology.

The minimum description length (MDL) principle \cite{10.1002/0471667196.ess1641.pub2} measures the representation of an object or a model by its smallest possible encoding size in bytes. This principle is used for model selection for predictive models, including regression \cite{doi:10.1198/016214501753168398} and classification \cite{Mehta1996}. MDL methods have also been applied for model selection of \textit{data representations} including clusterings, time series \cite{Hu:2011:DIC:2117684.2118274, doi:10.1198/016214501753168398}, networks \cite{doi:10.1093/bioinformatics/btl364} and network summarization \cite{doi:10.1137/1.9781611973440.11}. MDL has been used for structural outlier and change detection in dynamic networks \cite{Ferlez4497545, Sun:2007:GPM:1281192.1281266}. Our methodology encodes a collection of predictive models, together with the underlying network representation for model selection. No known work encodes both of these objects for model selection. 

Our work is orthogonal to graph kernels \cite{Vishwanathan:2010:GK:1756006.1859891} and graph embeddings \cite{2017arXiv170905584H}. These methods traverse network structure to extract path features and represent nodes in a low-dimensional space, or to compare nodes/graphs by shared structural features. We treat this graph traversal as one possible ordering of nodes for input to local tasks, but do not compare graphs or nodes directly; our focus is to evaluate how well the fixed model performs a relational learning task. 











\subsection{Contributions}

In this work, we formulate a task-focused model selection methodology for networks, and an extension for \textit{parsimonious} network model selection. 

\begin{compactitem}
    \item \textit{Model selection methodology}: We propose a generalized approach for evaluating networks inferred from data for performing particular tasks. 
    \item \textit{Network efficiency}: We propose a general minimum description length (MDL) \textit{efficiency} measure on the encoding of network representations and task models, which is comparable over varying models and datasets and encourages model sparsity.
    \item \textit{Validation and significance} We empirically demonstrate stability, sensitivity, and significance testing of models chosen by our model selection methodology.
\end{compactitem}

Our work focuses on process and methodology; the development of novel network inference methods and task methods are complementary but distinct from this work. Our work demonstrates that the network definition is a crucial step for evaluating predictive methods on networks. 


%

\section{Methods}

Model selection for task-focused network inference selects a network model and associated parameters of structured predictors to best perform a task or set of tasks on the topology. We ultimately evaluate the \textit{efficiency} of that coupling of the network and a method's performance on a given task.   

Figure \ref{fig:overview} presents a high-level overview of our model selection methodology. In Figure \ref{fig:overview}(a) as input, we are given raw data, assumed to be about a set of entities, which we will define as nodes, their attribute-set, and a node label-set.\footnote{For simplicity, we assume entities are unambiguously defined (e.g. users). Network inference methods which define nodes from raw data (e.g. sensor time series) follow this same methodology} Attributes are any data associated with nodes. These are often very high dimensional, for example, user activity logs and post content in social networks, gene expression profiles in gene regulatory networks, or full document text and other metadata in content networks. Labels are a notational convenience that allows differentiating input attributes from an attribute of interest for some predictive task (e.g. label inference, link prediction). Labels are typically low-cardinality fields (e.g. boolean, categorical).

In Figure \ref{fig:overview}(b), we apply a collection of network inference methods to generate separate network topologies. For example, we illustrate a threshold network, where edges are defined by a global similarity threshold between node attribute distributions. In contrast, the $k$-nearest neighbor network (KNN) defines edges as the $k$ most similar nodes to a given node. In Figure \ref{fig:overview}(c), for each inferred network, we apply several node weight functions that yield a subset of nodes to build a predictor. We illustrate network adjacency ({\color{moss}green}), and the community affiliation ({\color{midnight}blue}) of node $i$, where multiple colors represent nodes in both categories.  

In Figure \ref{fig:overview}(d), for each network, each of these weight functions yields a subset of nodes for the training of task predictors (e.g. node label inference, link prediction). Each row corresponds to a supervised classification task trained on attributes and labels in the node subset given by a different node weight rule. We evaluate all combinations of networks and their weighting functions and select a network inference-weight function pair. In our example, our methodology chooses a threshold network, where predictors are trained on network adjacency. 


\subsection{Preliminaries and notation}
Let $V=\{1...,i...,n\}$ be a set of nodes representing entities within the system. Let $A=\{\vec{a}_1, ..., \vec{a}_n\}$ be a set of attribute-vectors and $L=\{l_1, ..., l_n\}$ be a set of target labels of interest to be learned. Let $\mathcal{M}(A) \rightarrow E$ be a network inference function that yields an edge-set $E$ on some measure of node attributes. Let $\mathcal{W}_\mathcal{M}(i)\rightarrow U$ be a node weight function on the network inferred from $\mathcal{M}$ to produce a node subset $U$. Let $\classifier:A\times L \to L$ be a classifier trained on attributes and labels to produce a prediction $l'_i$ for node $i$.\footnote{Throughout, capital letters denote sets, lowercase letters denote instances and indices. Script characters (e.g. $\classifier$) and keywords (e.g. $\cost()$) denote functions, teletype (e.g. \knn-\bfs) denotes \textit{models}, square brackets (e.g. $A[U]$) denote the restriction of a set to a subset.}  %

\subsection{Node weighting/prioritization}

In our methodology, we induce a network structure using some method $\mathcal{M}$, (e.g. \knn{}). However, the topology of this network can be used in various ways to train a predictor for a given node $i$. We abstract this modeling choice as a \textit{node weight function}:

\begin{equation}
\networkquery_\mathcal{M}(i) \rightarrow U, \textrm{ where } U \subseteq V 
\label{eq:nodeorder}
\end{equation}

This function can be implemented as matrix $W$, where element $w_{i,j}$ is the probability that node $j$ is in $U$, where a weighted sampling on $\vec{w}_i$ yields $U$. This definition is general enough to encompass both deterministic and probabilistic functions. Deterministic functions include network adjacency. Probabilistic functions can be estimated using re-sampling techniques such as the bootstrap \cite{efron93bootstrap}. For example, we sample a node's community node-set uniformly at random to produce a node subset $U$ of fixed size $k$. Similarly, breadth-first node ordering from node $i$ is used to produces a fixed-length node subset of size $k$, where order at each depth is randomized. The above functions illustrate that $W$ need not only be the weighted network of $E$ (e.g. weighted network inference) but may incorporate higher-order such as communities or higher-order neighbors. 

Our general formulation allows us to define or learn any node weight function which outputs node subsets for training, including non-network functions not subject to edge-set $E$. In geometric space, $\networkquery_{l_2}(i)$ may order nodes by increasing Euclidean distance from $i$. We also define heuristics derived only from statistics on node attributes or network measures such as degree (see: Section \ref{subsubsec:nodeordering}) which measures the extent that a small number of exemplar nodes are predictive of the network.

Below, we focus on probabilistic functions that are comparable across node subsets of a fixed size $k$. For simplicity, we omit the bootstrapping notation since it is identical to the deterministic case for model selection.

\subsubsection{Network Model Selection}

 Let $\networkqueryr(i)\rightarrow U_i$ be a node weight function outputting node subsets $U_i$ for each node $i$. We train each classifier from the attributes and labels of these sampled nodes. Let $\classifierset$ be the resulting set of trained predictors on $r = \networkqueryr$: 
\begin{equation}
\begin{aligned}
\classifierset = \{\classifier(A[U_1], L[U_1]),...\\
\classifier(A[U_n], L[U_n])\}
\end{aligned}
\end{equation}

Let $\correctbase(\classifierset)$ be an evaluation measure over all predictors in $\classifierset$. In our case, we use \textit{precision} as the base measure because each of our applications is a rare-label classification task. 

We can now formally define our network model selection problem. This problem selects the (1) network inference function $\mathcal{M}$, and (2) the node weight function $\networkquery$ which maximizes `$\correctbase$':

\begin{algorithm2e}
  \SetAlgorithmName{Problem}
  \\
  \\
  \KwGiven{A set of nodes $V$, node attribute-set $A$, node label-set $L$, network inference functions $\mathcal{M} \in M$, node weight functions $\networkqueryr \in S$, Task predictor $\classifier(.)$}
  \KwFind{Node weight function $\networkquery_{\mathcal{M}}' \in S$}
  \KwWhere{$\networkquery_{\mathcal{M}}' = \argmax_{r} \correctbase(C_r)$}
  \caption{Task-Focused Network Inference Model Selection}
  \label{p:networkinference}
\end{algorithm2e}

This selected representation is the best for performing a particular task on the underlying nodes, e.g. link prediction or label prediction. In the case of our probabilistic weighting functions that sample at a fixed size of $k$, this is a hidden hyperparameter that yields distinct weight functions at each $k$. Selecting on these functions also selects the best (fixed) $k$. 

\subsection{Network Efficiency}
Our initial problem statement doesn't account for the representation size for performing the predictive task. For example, if a dense network and a sparse network have a similar precision on the task, the sparse network is a more \textit{parsimonious} model for the task. Therefore, we formulate a minimum description length (MDL) model selection criterion, which accounts for this representation cost.  

Let $\correct(\classifierset)$ simply be the sum of correct predictions over all predictors on test input. Analogous to precision above, correct predictions are appropriate for our class-imbalanced prediction problem and will yield an interpretable MDL measure. 

\begin{definition}
The {\em efficiency} of a network and its weighting $\networkqueryr$ for a node $i$ is the number of correct predictions per byte (higher is better). As above, we handle the hyperparameter $k$ in the probabilistic case which cancels out of the deterministic definition. The efficiency $\mathcal{E}$ of an individual node is defined:
\begin{equation}
\mathcal{E}(\networkqueryr, i) = \max_k\left\{{\frac{\correct(\classifierset[i,k])}{\sum\cost(\classifierset[i,k])}}\right\}. 
\label{eq:efficiency}
\end{equation}
\end{definition}
where $\cost(.)$ is the representation cost of encoding the predictor(s) for node $i$.

Let $\kappa_i=\argmax_{k}\mathcal{E}(\networkqueryr, i)$, the $k$ associated with the maximum efficiency of node $i$. We then sum over these $\kappa$ to get the most efficient (variable-sized $k$) representation over all nodes:



\begin{equation}
\begin{split}
\taskcorrect{} = \sum_i\correct(\classifierr [i, \kappa_i])\\
\taskcost{} =\sum_i\cost(\classifierr [i,\kappa_i])
\end{split}
\label{eq:efficiencyseparate}
\end{equation}
Then, the overall efficiency of $\networkqueryr$ is given by:
\begin{equation}
\efficiency = \frac{\taskcorrect}{\taskcost + \netcost}. 
\label{eq:efficiencyfull}
\end{equation}

We can then use this definition to re-formulate an MDL criterion of our above problem statement:

\begin{algorithm2e}
  \SetAlgorithmName{Problem}
  \\
  \\
  \KwGiven{A set of nodes $V$, node attribute-set $A$, node label-set $L$, 
  network weight function-set $S$ where $\networkqueryr \in S$, Task $\classifier(.)$}
  \KwFind{Network weight function $\networkquery_{\mathcal{M}}' \in S$}
  \KwWhere{$\networkquery_{\mathcal{M}}' = \argmax \efficiency$}
  \caption{MDL Task-Focused Network Inference Model Selection}
  \label{p:networkinference_mdl}
\end{algorithm2e}

For brevity, we refer to `model selection' simply as selecting the network representation and its node weight function. A key point is that we measure the efficiency of the \textit{node weighting function}. In many instances $\networkqueryr$ may be implemented more with greater space-efficiency than the edge-set produced by $\mathcal{M}$. For example, community/cluster affiliation can be encoded as a single list of size $\bigo(|V|)$. Therefore, the model selection on our efficiency criterion measures whether a network is a better model than other non-network weight functions, group-wise weighting (e.g. communities), etc. accounting for the varying representation cost of each.

Previous work has evaluated network model-selection for robustness to multiple tasks (e.g. link prediction, label prediction) as well as different base predictor models (e.g. random forests, support vector machines) \cite{brugereicdm2017}. Problem \ref{p:networkinference_mdl} can straightforwardly select over varying underlying network definitions, tasks, or task predictors that maximize efficiency. We simplify the selection criteria to focus on measuring the efficiency of node weight functions and their underlying representations, but this current work is complimentary to evaluating over larger parameter-spaces.

\subsection{Network Models}

We define several network models for our study. Our focus is not to propose a novel network inference model from attributes (see: \cite{Pfeiffer:2014:AGM:2566486.2567993,Kim2012}). Instead, we apply existing common and interpretable network models to demonstrate our framework. The \textit{efficiency} of any novel network inference model should be evaluated against these standard baselines. 

A network model $\mathcal{M}$ constructs an edge-set from node attributes and/or labels: 
\begin{equation}
\mathcal{M}_j: \mathcal{M}_j(A,L) \rightarrow E_j. 
\end{equation}
We use $k$-nearest neighbor (\knn) and Threshold (\tth) models which are ubiquitous in many domains. 

Given a similarity measure $\mathrm{d}(\vec{a}_i, \vec{a}_j) \rightarrow s_{ij}$ and a target edge count $\rho$, this similarity is used to produces a pairwise attribute similarity space. We select edges by:


\begin{compactitem}
	\item $k$-nearest neighbor $\mathcal{M}_{\knn}(A, \mathrm{d}(.), \rho)$: for a given $i$, select the top $\lfloor{\frac{\rho}{|V|}}\rfloor$ most similar $\mathrm{d}(\vec{a}_i, \{A \setminus \vec{a}_i\})$. In directed networks, this produces a network which is $k$-regular in out-degree, with $k=\lfloor{\frac{\rho}{|V|}}\rfloor$.
	\item Threshold $\mathcal{M}_{\tth}(A, \mathrm{d}(.), \rho)$: over all pairs $(i,j)\in V$ select the top $\rho$ most similar $\mathrm{d}(\vec{a}_i, \vec{a}_j)$.
\end{compactitem} 

Let these edge-sets be denoted $E_{\knn}$ and $E_{\tth}$, respectively. We use varying network sparsity ($\rho$) on these network models to define `sparse' or `dense' network models. Similarity measures may vary greatly by the application. We use cosine similarity, which is ubiquitous, especially in information retrieval and recommender system applications.

\subsection{Node Weight Functions}
\label{subsubsec:nodeordering}
Table \ref{tab:functions} summarizes the node weight functions (Equation \ref{eq:nodeorder}) used in our methodology. We define only a small number of possible functions, focusing on an interpretable set that helps characterize the underlying system. The encoding cost increases from top to bottom.\footnote{For simplicity, we refer to models only by their subscript labels, e.g. \knns-\bfs{} representing breadth-first search weighting on the sparse KNN network.}

\subsubsection{Importance Weighting Heuristics}

The \activityflat{} and \degreeflat{} heuristics are the simplest weighting functions we define. \activityflat{} uses importance sampling with weights proportional to the number of non-zero attributes of a node. Similarly, we define \degreeflat{} with importance sampling on the node degree distribution. Although \degreeflat{} uses a network measure to determine the degree ranking, we need only encode a list of nodes to sample, so the encoding cost is $\bigo(|V|)$ space. 

\subsubsection{Community, Random Heuristics}

The \cluster{} weighting method applies Louvain graph clustering \cite{ICT4DBibliography2429} on the input edge-set $E$. This reduces the network to a $|V|$-length list representing the community assignment of each node. Although \cluster{} is derived from an underlying network, we only encode the reduced representation. This measures whether the network is better represented as a collection of \textit{groups} where individual edges are uninformative for the predictive task, in $\bigo(|V|)$ space rather than $\bigo(|E|)$ space. The \random{} weighting yields unbiased, random subsets on V. For this, we need only encode all node IDs, in $\bigo(|V|)$ space.

\subsubsection{Graph-traversal Weighting (Local)}

The \bfs{} weighting method uses a breadth-first exploration on an underlying edge-set $E$ from seed $i$. This is encoded by an adjacency list in $\bigo(|E|)$. In other analyses, we use either fixed neighborhood weighting or egonet weighting (in the link-prediction task). Each is a special case of \bfs{}. We also refer to these as `local' node weight methods.

The \activity{} weighting method is an inferred network where all out-edges point to some node in the top-$\ell$ ($=0.1$) fraction of nodes with respect to \activityflat{}. This is a \knn{} network, constrained to high-activity nodes. The additional encoding cost relative to \activityflat{} measures the extent that specialization exists in the top-ranked individuals as a set of exemplars for all nodes (i.e. the individual exemplar relations matter). \degree{} is defined analogously with respect to \degreeflat{}. 

\begin{table}
\centering
\begin{tabular}{|l|l|}
\hline
Node Weight Method & Encoding Cost\\
\hline \hline
$\networkquery_{\activityflat}$  & $\bigo(|V|)$\\
\hline
$\networkquery_{\degreeflat}$  & $\bigo(|V|)$\\
\hline
$\networkquery_{\cluster}$ & $\bigo(|V|)$ \\
\hline
$\networkquery_{\random}$  & $\bigo(|V|)$\\
\hline
\hline
$\networkquery_{\bfs}$ & $\bigo(|E|)$ \\
\hline
$\networkquery_{\activity}$  & $\bigo(|V|\times \ell)$\\
\hline
$\networkquery_{\degree}$  & $\bigo(|V|\times \ell)$\\
\hline
\end{tabular}
\caption{The input signature and space complexity of node weight functions. We define four functions implemented by non-network structures (top), and three implemented by a network (bottom)}
\label{tab:functions}
\end{table}

\subsection{Tasks for Evaluating Network Models}

We evaluate network models on two fundamental network tasks: collective classification and link prediction. 

\subsubsection{Collective Classification (CC)} The collective classification problem learns relationships between network edge structure and attributes and/or labels to predict label values \cite{sen:aimag08, 4476695}. This task is often used to infer unknown labels on the network from `local' discriminative relationships in attributes, e.g. labeling political affiliations, brand or media preferences.


\subsubsection{Link Prediction (LP)}

The link prediction problem \cite{Liben-Nowell2007} learns a method for the appearance of edges from one edge-set to another. Link prediction methods can incorporate attribute and/or label data, or using simple structural ranking \cite{Adamic2003}.

For this problem, we fix the weighting method to generate node pairs within the `egonet' induced subgraph. We construct edge attributes by the difference in node attribute vectors and learn a balanced edge/non-edge predictor on these features.

\subsubsection{Task Predictors}

To demonstrate our methodology, we use Random Forest and Support Vector Machine (SVM) task predictors. The Random Forest is straightforwardly encoded as its underlying decision trees (lists), and the aggregation weights of the trees. SVMs are similarly encoded as the hyperplane coefficients.

\subsection{Measuring Efficiency}

Our node weight functions and task predictors can now be represented as a byte-encoded object `o' (e.g. lists). We can now measure efficiency (Equation \ref{eq:efficiencyfull}). To implement our $\cost(o)$ function estimating the minimum description length, we convert each object representation to a byte-string using LZ4 compression,\footnote{\url{https://lz4.github.io/lz4/}} (analogous to zip) which uses Huffman coding to reduce the cost of the most common elements in the representation string (e.g high degree nodes, frequent features in the random forest). This method was chosen primarily for runtime performance since we do many compression operations over each candidate network topology. Finally, we report the length of the compressed byte-string:
\begin{equation}
    \cost(o) = |\mathrm{lz4.dumps}(\mathrm{json.dumps}(o))|	
\end{equation}
\subsubsection{Node weight function reach}
\label{subsubsec:reach}
Let the reach-set of $C_r$, $\mathrm{reach}(C_r)$ be defined as the set of nodes accessed at least \textit{once} to train \textit{any} predictor in $C_r$.  We define $\networkquery^{*}_r$ as the representation of $\networkquery_r$ including only nodes in the reach-set. If the underlying representation of $\networkquery_r$ is a graph, the $\networkquery^{*}_r$ representation is an induced subgraph where \textit{both} nodes incident to an edge are in the reach-set. If $\networkquery_r$ is represented as a list, we simply remove elements that are not in the reach-set.

The encoding size $\cost(\networkquery^*_r)$ measures only the underlying representation accessed for the creation of the $C_r$ task predictor set. This is a more appropriate measure of the representation cost. In practice, $|\mathrm{reach}(C_r)| << |V|$. For our evaluation, we always report $\mathcal{E}(\networkquery^*_r)$.

\section{Datasets}

We demonstrate our model selection methodology on label prediction tasks of three different online user activity datasets with high-dimensional attributes: beer review history from \ba{}, music listening history from \lfm{}, and movie rating history from \ml{}. 

\small{}
 \begin{table}[t]
 	\centering
 	\resizebox{\boxscale\columnwidth}{!}{
	\begin{tabular}{|c|c|c|c|c|}
		\hline
		Dataset & $|V|$ & $|A|$ & Labels & $|L|$\\\hline
		\lfm{} 20K \cite{mlg2017_13} & 19,990 &  1.2B & 8 & 16628\\
		\ml{} \cite{Harper:2015:MDH:2866565.2827872} & 138,493 & 20M &8&43179 \\ 
		\ba{} \cite{McAuley:2012:LAA:2471881.2472547} & 33,387 & 1.5M &8 & 13079  \\
		\hline
	\end{tabular}}
 	\caption{A summary of datasets in this paper. $|L|$ reports the total number of positive node labels over 8 labelsets.} 
 	\label{tab:data}
 \end{table}
\normalsize{}

\subsection{\lfm{}}
\lfm{} is a social network focused on music listening, logging and recommendation. Previous work collected the entirety of the social network and associated listening history, comparing the social network to alternative network models for music genre label prediction \cite{mlg2017_13}.

Sparse attribute vectors $\vec{a}_i\in A$ correspond to counts of artist plays, where a non-zero element is the number of times user $i$ has played a particular unique artist. \lfm{} also has an explicit `friendship' network declared by users. We treat this as another possible network model, denoted as $E_{\social}$, and evaluate it against others. 

We evaluate the efficiency of node weight functions for the label classification task of predicting whether user `$i$' is a listener of a particular genre. A user is a `listener' of an artist if they have at least 5 plays of that artist. A user is a `listener' of a genre if they are a listener of at least 5 artists in the top-1000 most tagged artists with that genre tag, provided by users. We select a subset of 8 of these genre labels (e.g. `dub', `country', `piano'), chosen by the guidance of label informativeness from previous work \cite{mlg2017_13}. 
\subsection{\ml{}} 

\ml{} is a movie review website and recommendation engine. The \ml{} dataset \cite{Harper:2015:MDH:2866565.2827872} contains 20M numeric scores (1-5 stars) over 138K users. 

Sparse non-zero attribute values correspond to a user's ratings of unique films. We select the most frequent user-generated tag data thatcorresponds to a variety of mood, genre, or other criteria of user interest (e.g. `inspirational', `anime', `based on a book'). We select 8 tags based on decreasing prevalence (i.e. `horror', `musical', `Disney'), and predict whether a user is a `viewer' of films of this tag (defined similarly to \lfm{} listenership).

\subsection{\ba{}} 

\ba{} is a website containing text reviews and numerical scores of beers by users. Each beer is associated with a category label (e.g. `American Porter', `Hefeweizen'). We select 8 categories according to the decreasing prevalence of the category label. We predict whether the user is a `reviewer' of a certain category of beer (defined similarly to \lfm{} listenership).

\subsection{Experiment Set-up}

When building \knn~and \tth~networks, we construct both `dense' and `sparse' models, according to edge threshold $\rho$. For \lfm{}, we fix $\rho= |E_{\social}|$ for the dense network, and $\rho= 0.5 \times |E_{\social}|$ for the sparse. For both \ba{} and \ml{}, a network density of $0.01$ represents a `dense' network, and $0.0025$ a `sparse'. However, all of these networks are still `sparse' by typical definitions. 

User labels on all three datasets are binary (the user is a listener/reviewer of this genre/category), and sparse. Therefore, we use a label `oracle' and present only positive-label classification problems. This allows us to evaluate only distinguishing listeners etc. rather than learning null-label classifiers where label majority is always a good baseline. 

Table \ref{tab:data} ($|L|$ columns) reports the count of non-zero labels over all 8 label-sets. This is the total number of nodes on which evaluation was performed. Each  task predictor can be independently trained and evaluated, allowing us to scale arbitrarily. Following \citet{brugereicdm2017}, for each of the three datasets, we split data into temporally contiguous `validation,' `training,' and `testing,' intervals of approximately 1/3 of each dataset. Training is on the middle third, validation is the interval prior, and testing on the latter interval. Model selection is performed on validation, and this model is evaluated on testing. 

Below, we perform two primary empirical studies. First, we study network model selection in the presence of multiple tasks. We demonstrate that different network definitions and node weight functions are selected per task. Furthermore, jointly selecting for multiple tasks may not perform well for either. Second, we measure the \textit{efficiency} of selected models and demonstrate that tasks can often be more parsimoniously represented by simpler, non-network representations.
\section{Empirical analysis: Consistency and Multiple Tasks}

We validate network models under varying hyperparameters on each dataset. We measure \textit{precision} on both collective classification (CC) and link prediction (LP). For each network model (i.e. network representation, node weight method and associated hyperparameters), we rank models on the precision of the induced predictors over all nodes in validation and \textit{select} the top-ranked model for evaluation in testing. To evaluate the robustness of model selection, we compare all selected network models on both validation and testing partitions to examine their full ranking.
 
\subsection{Model Stability: Precision}
\label{subsec:performance}

\begin{table}
	\centering
	\resizebox{\boxscale\columnwidth}{!}{
	\begin{tabular}{|cccc|cc|}
\hline
\multicolumn{4}{|c|}{Precision (Testing)} & \multicolumn{2}{c|}{Validation vs. Testing} \\
\hline
\multicolumn{1}{|c}{Task-Predictor} &$\mu$ & $\mu_{(\topk)}$  & $p_{(1)}$ & $\Delta p_{(1)}$ & $\mathrm{rank}$ \\
\hline \hline
\multicolumn{6}{|c|}{\textbf{BeerAdvocate}} \\
 \hline

CC-RF & 0.12 &  0.20 & 0.23 &  \textbf{-0.01} & \textbf{0.99}\\
CC-SVM & 0.35& 	0.64 & 0.70 & \textbf{-0.03} & \textbf{0.99} \\
\rowcolor{black!20}
LP-RF & 0.50 & 0.53 & 0.58 &  -0.11 & 0.09 \\
\rowcolor{black!20}
LP-SVM &0.51 &	0.57& 0.64 &  -0.04 & \textbf{0.99}\\

\hline

\multicolumn{6}{|c|}{\textbf{Last.fm}} \\
 \hline

       CC-RF &  0.18 &       0.38 & 0.39    &               \textbf{-0.01} & \textbf{0.90} \\
       CC-SVM &  0.38 &       0.62 &  0.64   &                 \textbf{-0.01} & \textbf{0.91}\\
\rowcolor{black!20}
      LP-SVM &  0.53 &       0.60 & 0.68    &              \textbf{-0.00}& \textbf{1.00}\\
\hline

\multicolumn{6}{|c|}{\textbf{MovieLens: Genres}} \\
 \hline

       CC-RF &  0.01 &       0.03 & 0.06    &               -0.05 & 0.61\\
       CC-SVM &  0.15 &       0.30 & 0.36    &              \textbf{-0.01} & \textbf{0.95}\\
\rowcolor{black!20}
       LP-RF &  0.46 &      0.48 & 0.6    &                -0.21& 0.06\\
\rowcolor{black!20}
      LP-SVM &  0.45 &      0.47 &  0.52   &               -0.09 & 0.29\\
\hline
\multicolumn{6}{|c|}{\textbf{MovieLens: Tags}} \\
 \hline
CC-RF & 0.28 & 0.60 & 0.68 & -0.10 & \textbf{0.92} \\
CC-SVM & 0.55 &	0.80 & 0.86 &-0.06 & 0.87\\
\hline
\end{tabular}

}
	\caption{Task precision over all datasets and methods. $\mu$ reports the mean precision over all models in test. $\mu_{(\topk{})}$ reports the mean precision over the top-\topk{} models in test, and $p_{(1)}$ the precision of the top-ranked model. $\Delta p_{(1)}= p_{s} - p_{(1)} $ reports the precision of the selected model $s$ vs. the best model both evaluated in test (0 is best). $\mathrm{rank}$ reports the percentile ranking of the selected model, as evaluated in test (1 is best). Link prediction tasks (grey rows) are balanced tasks with a random baseline of $0.5$. \textbf{Bold} indicates significant values: $\frac{\Delta p_{(1)}}{p_{(1)} - \mu} \leq 0.05$.}
	\label{tab:perf}
\end{table}

Let $p_i$ denote the precision of the $i$-th model, $p_s$ as the precision of the selected model evaluated on the testing partition, $p_{(1)}$ as the precision of the `best' model in test. Let $\mu$ be the mean precision over all models ($|N|> 100$) and $\mu_{(\topk{})}$ the mean precision over the top-$\topk{}$ models.

Table \ref{tab:perf} summarizes performance over different tasks and predictors. We report the mean precision of the top-10 ranked models ($\mu_{(\topk{})}$) vs. the precision of the top-ranked model ($p_{(1)}$),  both evaluated in test.

The best model in validation need not be the best possible model in testing. Table \ref{tab:perf} reports $\Delta p_{(1)}= p_{s} - p_{(1)}$, the difference in precision between the selected model $s$, and the best possible model, both evaluated in test (0 is best). We use bold to indicate significant values: $\frac{\Delta p_{(1)}}{p_{(1)} - \mu} \leq 0.05$, i.e. a change less than a $0.05$ factor of the maximum lift vs. mean precision in test. Finally, $\mathrm{rank}$ reports the percentile ranking of the selected model, as evaluated in test (1 is best). 

In Table \ref{tab:perf}, $\Delta p_{(1)}$ and $\mathrm{rank}$ are both measures of model consistency. Selected models often approximate the best model in test, and have a high rank among other models. Selected models with high error to the best test model can have arbitrarily bad ranking in test (e.g. LP-RF in BeerAdvocate). Below, we look more closely at the selected model, as well as stability in deeper model rankings.

\subsection{Model Consistency: Selected Model Ranking}

\addtolength{\tabcolsep}{-3pt}   
\begin{table}[ht]
	\centering
	\resizebox{\boxscale\columnwidth}{!}{

\begin{tabular}{|cc|cc|}
\hline
\multicolumn{4}{|c|}{Selected Network-Weighting Function vs. $\mathrm{rank}$} \\
\hline
 CC-RF & CC-SVM & LP-RF & LP-SVM \\
\hline \hline
\multicolumn{4}{|c|}{BeerAdvocate}\\
\hline
\textbf{0.99} & \textbf{0.99} & 0.09 & \textbf{0.99} \\
\textbf{KNN-Local}& \textbf{TH-Local} & KNN-Activity & TH-Activity\\
\hline
\multicolumn{4}{|c|}{Last.fm}\\
\hline
\textbf{0.90} & \textbf{0.91} & -- & \textbf{1.00} \\
\textbf{KNN-Community} &    \textbf{Social-Community} &   --&        \textbf{TH-Local} \\
\hline
\multicolumn{4}{|c|}{MovieLens: Genres}\\
\hline
0.61 & \textbf{0.95} & 0.06 & 0.29 \\
KNN-Local &           \textbf{TH-Community} &           TH-Activity & TH-Activity \\
\hline
\multicolumn{4}{|c|}{MovieLens: Tags}\\
\hline
\textbf{0.92} & 0.87 & -- & -- \\
KNN-Local &         KNN-Local &    -- & -- \\
\hline

\end{tabular}

	}
	\label{tab:results_rank}
	\caption{The normalized rank of the selected model, evaluated in test. $\mathrm{rank}=1$ indicates the best models in both validation and testing are the same. The row below the $\mathrm{rank}$ indicates the selected network and node weight function. \textbf{Bold} indicates all selected models with $\mathrm{rank} \geq 0.9$, i.e. in the top $10\%$ of models.}
	\label{tab:rank_diff}
\end{table}
\addtolength{\tabcolsep}{3pt}   

Table \ref{tab:rank_diff} reports the normalized $\mathrm{rank}$ of the selected model, evaluated on test as reported in Table \ref{tab:perf}. We highlight models with high rank-consistency between validation and test: $\mathrm{rank} > 0.9$, i.e. the selected model is in the top $10\%$ of models in test.

Table \ref{tab:rank_diff} shows several cases of high rank-inconsistency (e.g. BeerAdvocate LP-RF, MovieLens LP-RF) and high consistency for others (BeerAdvocate CC-RF, CC-SVM). These results are over many ranked network models ($|N| > 100$). In Last.fm, community weighting is selected for both CC methods. The Social-Community model is selected for CC-SVM. For BeerAdvocate on CC, local models are consistently selected and have a high rank in testing, even though SVM and RF methods have very different performance in absolute precision. This table demonstrates that the appropriate network model changes according to not only the dataset, but different tasks on the same dataset.

\subsection{Model Stability: Rank Order}

\begin{table}[htbp]
	\centering
	\resizebox{\boxscale\columnwidth}{!}{
	\begin{tabular}{|c|ccc||c|}
\hline
\multicolumn{5}{|c|}{Rank Ordering (Validation vs. Test)} \\
\hline
 \multicolumn{1}{|c}{Task-Predictor}  & $\tau$ & $p$-value &  $\mathrm{intersection}_{(10)} $ & \textbf{Total} \\
\hline \hline
\multicolumn{5}{|c|}{BeerAdvocate}\\
\hline
CC-RF & 0.7   &\textbf{1.75E-34} & \textbf{6} & 4 \\
CC-SVM & 0.6  &\textbf{1.54E-25} & \textbf{8} & 4 \\
LP-RF & 0.34  & \textbf{3.08E-09} & 3  & 1 \\
LP-SVM & 0.44 &\textbf{1.09E-14} & \textbf{5} & 3  \\
\hline
\multicolumn{5}{|c|}{Last.fm}\\
\hline
CC-RF &   0.88 & \textbf{2.35E-24} & \textbf{9} & 4 \\
CC-SVM &  0.88 & \textbf{5.85E-21} & \textbf{8} & 4 \\
LP-SVM &   0.70& \textbf{8.15E-14} & \textbf{8} & 4 \\ 
\hline
\multicolumn{5}{|c|}{MovieLens: Genres}\\
\hline
CC-RF &   0.15 & 6.89E-03          & 0 & 0\\
CC-SVM &  0.57 & \textbf{9.99E-24} & 4 & 3\\
LP-RF &  -0.07 & 2.29E-01          & 0 & 0\\
LP-SVM & -0.07 & 2.40E-01          & 0 & 0\\
\hline
\multicolumn{5}{|c|}{MovieLens: Tags}\\
\hline
CC-RF & 0.61 & \textbf{8.95E-27} & 0 & 2\\
CC-SVM &0.52 & \textbf{4.43E-20} & 1 & 1\\

\hline
\end{tabular}

}
	\label{tab:results_full}
	\caption{The Kendall's $\tau$ precision rank order correlation between models in validation and test. 1 indicates the rankings are the same, 0 indicates random relative ordering. $\tau_{\topk{}}$ reports rank order correlation on the top-\topk{} models. We report associated p-values. \textbf{Bold} indicates the models with $p <0.001$ and $\mathrm{intersection}_{10} \geq 5$.}
	\label{tab:tau}
\end{table}

Table \ref{tab:tau} reports the Kendall's $\tau$ rank order statistic between the precision ranking of models, for validation and testing. $\tau=1$ indicates the rankings are identical. We report the associated $p$-value of the $\tau$ rank order statistic. For most CC tasks, model ranking is particularly consistent. Similarly, both tasks on \lfm{} have high rank-correlation. 

While this ranking shows remarkable consistency, it's not suitable when the result contains many bad models, which may be arbitrarily placed at low ranks. To handle this, we report $\mathrm{intersection}_{(10)}$,  the shared models in the top-10 of validation and test. Since top-$k$ lists may be short and have disjoint elements, we find the simple intersection rather than rank order. We highlight tasks in bold at a rank order significance level of $p < 1.00\mathrm{E}{-03}$, and $\mathrm{intersection}_{(10)} \geq 5$.

Table \ref{tab:tau} `Total' summarize the count of bold entries across Tables \ref{tab:perf}, \ref{tab:rank_diff}, and \ref{tab:tau}. This corresponds to scoring models on several consistency criteria: (1) precision (2) selected ranking, (3) rank correlation, and (4) top-$10$ rank intersection.

MovieLens under `tag' labels reports a peculiar result. It performs very well at both $\mu_{(10)}$ and $p_{(1)}$ for both SVM and RF. However, it has a high $\Delta p_{(1)}$ and low $\mathrm{intersection}_{(10)}$. Looking closer at the results, two similar groups of local models perform well. However, in validation, this is under an adjacency local model, and the testing partition favors a wider BFS local model. However, both of these models are similar; this similarity isn't reflected in our ranking.  


\subsection{Consistency: Node Weight Functions}

\begin{figure}[htbp]
\centering
\includegraphics[width=\figwidth\columnwidth]{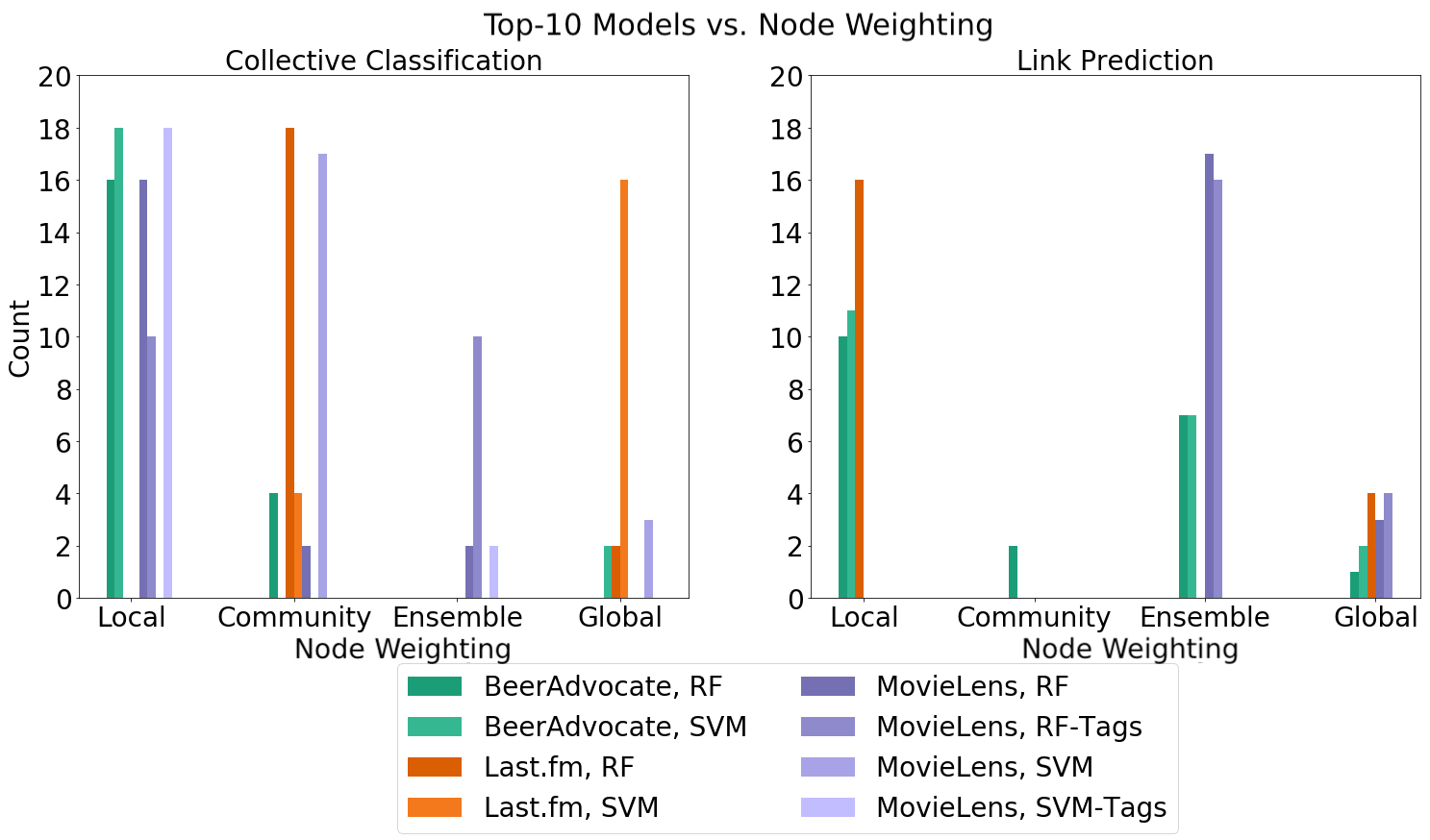}
\caption{Counts of node weight functions of the top-\topk{} ranked models in validation and testing (\twotopk{} total). Primary colors denote different datasets, shades denote different task predictors.}
\label{fig:locality}
\end{figure}

Our framework allows further investigation of weighting functions suitable for particular types of tasks, measured by their ranking. Figure \ref{fig:locality} reports the counts of top-\topk{} in validation and test, grouped by their weighting function, on CC (left) and LP (right) tasks. Each principal color represents a dataset, and shades denote different task predictors. 

For CC, BeerAdvocate favors local task models, while Last.fm favors community and global models; both of these results agree with model selections in Table \ref{tab:rank_diff}. Global weighting measures the extent that population-level models (e.g. unbiased random population sampling) are favored to any structured weighting. Looking closer at Last.fm, the $\Delta p_{(1)}$ for the best-ranked Global model in test is only -0.01 for CC-SVM, and -0.05 for CC-RF. This indicates a weak network effect on Last.fm for CC under our evaluated models.

For each dataset, preferred models differ greatly by the task. LP on BeerAdvocate has an increased preference for activity weighting compared to CC. The preference for global weighting largely disappears for LP on Last.fm in favor of local models. This further demonstrates that we find representations better suited for different tasks, which change both by dataset and task.

\begin{figure}[htbp]
\centering
\includegraphics[width=\figwidth\columnwidth]{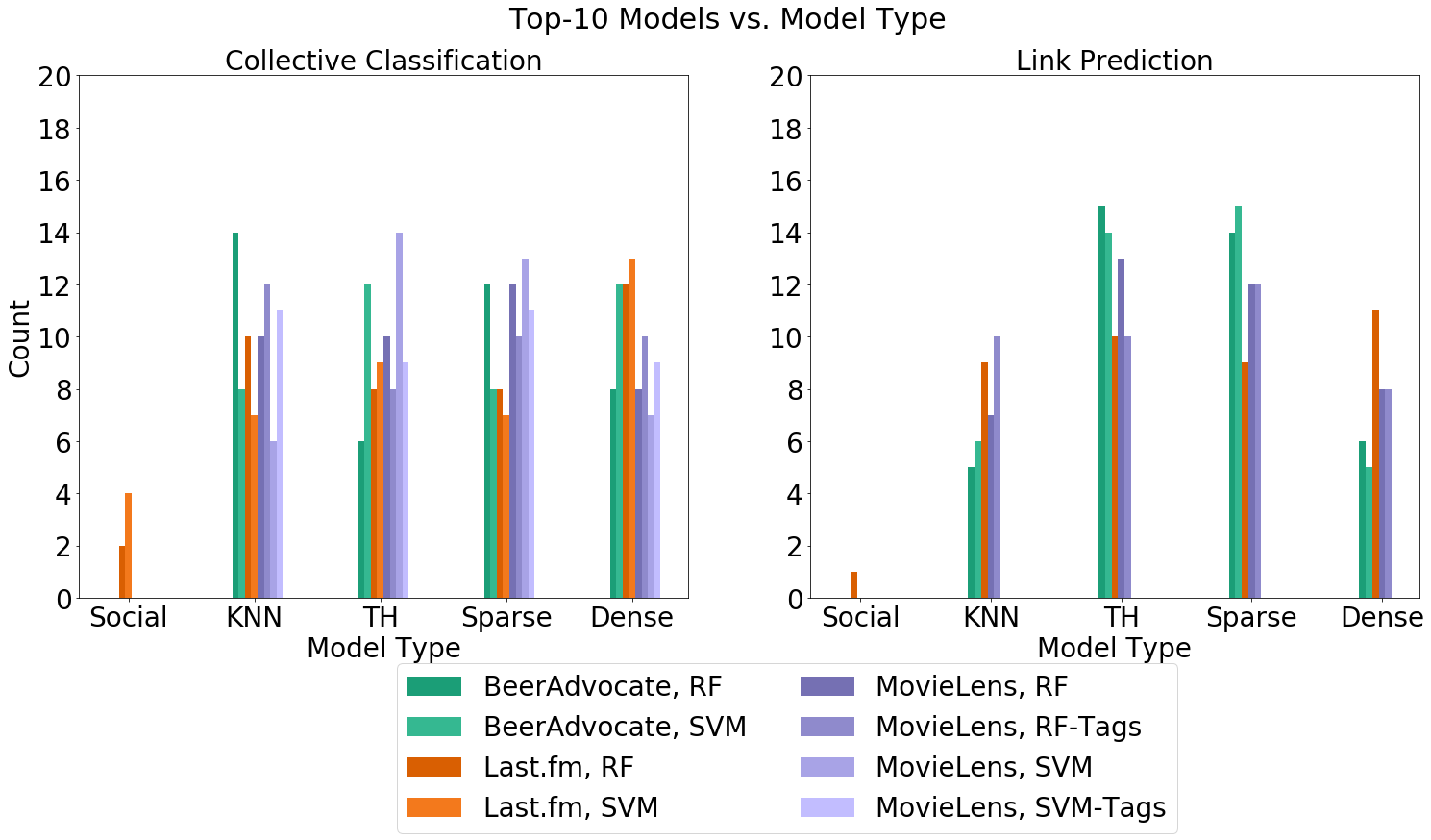}
\caption{Counts of network models and density associated with the top-\topk{} ranked models in validation and test (\twotopk{} total). Primary colors denote different datasets, shades denote different task methods.}
\label{fig:modeltype}
\end{figure}

Figure \ref{fig:modeltype} reports the counts of models according to underlying representation for the top-\topk{} models in validation and test. The first three bar groups report a total of \twotopk{} model configurations over `Social' (only Last.fm), `KNN,' and `TH' network models. The next two bar groups report \twotopk{} configurations over `Sparse' and `Dense' settings. `Sparse' refers to very sparse networks on the order of $0.0025$ density, while `dense' is on the order of densities observed in social networks (e.g. $0.01$).

For all tested datasets, there is not a strong preference for a particular network model or density for either CC or LP. However, this does not mean that precision is `random' over varying network models. The $\tau$ rank order and $\mathrm{intersection}_{10}$ are very consistent in several of these task instances (Table \ref{tab:tau}). Instead, node weight function preferences seem to drive the three datasets we examine, where the network representation will perform more similarly under the same weighting method than across different weighting methods. 

\begin{table}
	\centering
	\scalebox{\boxscale}{
	\begin{tabular}{|c|cc|cc|ccc|}
\hline
\multicolumn{8}{|c|}{Median $p_i - p_j$, match vs. mismatch (SVM)} \\
 \multicolumn{1}{|c}{} &  \multicolumn{2}{c|}{BeerAdvocate}  & \multicolumn{2}{c|}{Last.fm} & \multicolumn{3}{c|}{MovieLens} \\
\multicolumn{1}{|c}{} & \multicolumn{1}{c}{CC}  & LP & CC & LP & CC & CC-Tags & LP \\
\hline \hline

Weighting & -0.03 & 0.01 & -0.14 & -0.04 & -0.04 & 0.01 & -0.02 \\
Network & 0.00 & 0.00 & 0.01 & 0.01 & 0.00 & 0.00 & 0.00 \\
\hline
\end{tabular}
}
	\caption{The difference between medians of pairwise precision comparisons grouped by matching or mismatching criteria. More negative values denote higher median difference among mismatching models.}
	\label{tab:matchmismatch}
\end{table}

We directly evaluate this hypothesis in Table \ref{tab:matchmismatch}. We report the median of pairwise differences of precision between model pairs, by matched/mismatched network representation or node weighting: $\mathrm{median}(p_{i} - p_{j}) - \mathrm{median}(p_{k} - p_{l})$, where $(i,j)$ are all matched pairs grouped by the same network or weighting, and $(k,l)$ all mismatched pairs. More negative values represent higher differences in precision on mismatches than matches, for that row's criteria. Mismatching node weight functions indeed account for more difference in precision than mismatching network representations.

\subsection{Model Selection and Cross-Task Performance}

\begin{table}
	\centering
	\resizebox{\boxscale\columnwidth}{!}{
	\begin{tabular}{|c|cc|cc|}
\hline
\multicolumn{5}{|c|}{Cross-Task Model Ranking (SVM)} \\
\hline
 \multirow{2}{*}{Model Selection \textbackslash~Testing}  & \multicolumn{2}{c|}{CC-SVM} & \multicolumn{2}{c|}{LP-SVM} \\
 &\multicolumn{1}{c}{$\Delta p_{(1)}$}&$\mathrm{rank}$ & \multicolumn{1}{c}{$\Delta p_{(1)}$}&$\mathrm{rank}$\\
\hline \hline

BeerAdvocate, CC-SVM & -0.03&	0.99&	-0.17 &0.14\\
Last.fm, CC-SVM&-0.01 &	0.91&-0.16&0.68\\
MovieLens, CC-SVM& -0.01	&0.95 &	-0.08 &0.47\\
\hline
BeerAdvocate, LP-SVM & -0.65&	0.09 &-0.04 &	0.99 \\
Last.fm, LP-SVM	&-0.43& 0.21 &-0.00&1.00 \\
MovieLens, LP-SVM &-0.31&0.76 &-0.09&	0.29\\
\hline \hline
\multicolumn{5}{|c|}{Average-Precision Model Selection (SVM)} \\
\hline 
BeerAdvocate, SVM & -0.29&	0.60 &-0.10 &	0.84 \\
Last.fm, SVM	&-0.01& 0.94 &-0.13&0.75 \\
MovieLens, SVM &-0.01&0.98 &-0.09&	0.31\\
\hline

\end{tabular}

}
	\caption{(Upper) Performance of network models selected in validation (left), evaluated on varying tasks (top) according to the difference against the best model in test: $\Delta p_{(1)}= p_{s}-p_{(1)}$, for $s$ the selected model (0 is best). $\mathrm{rank}$ reports the rank of the selected model in test (higher is better). Diagonal entries correspond to models selected and evaluated on the same task (i.e. values from \ref{tab:perf}, and \ref{tab:rank_diff}), Off-diagonal correspond to models selected and evaluated on different tasks. (Lower) Average-Precision Model Selection using the average of precision on CC and LP.}
	\label{tab:cross}
\end{table}

We now aim to measure the performance of models across different tasks, and whether the same selected model tends to perform both tasks. Table \ref{tab:cross} (Upper) reports model performance when model selection and evaluation are on matching/mismatching tasks. We do model selection on validation (for each task on the left) and report task performance in testing, on the task method given by the column. The $\Delta p_{(1)}$  and $\mathrm{rank}$ are calculated as previously, where values on the diagonal are the same as in Tables \ref{tab:perf} and \ref{tab:rank_diff}, respectively. On the off-diagonal, the model is selected in validation on a different task than it is evaluated on in test. Table \ref{tab:cross} (Lower) reports model performance by doing model selection on the \textit{average} of CC and LP precision.

This result clearly demonstrates the main takeaway of this section: the `best' network model depends on the subsequent task. The off-diagonal shows that in every case, models selected on the `other' task perform very poorly. Consider the worst case for same-task selection--LP on MovieLens--scored 0 on our 4 selection criteria (Table \ref{tab:tau}), yet this selected model performs 3x better in $\Delta p_{(1)}$ than it's cross-task selected model. Over our three datasets, the \textit{average} factor increase in $\Delta p_{(1)}$ performance from model selection using same-task versus cross-task is $\approx 10x$. 

Perhaps we can simply do model selection jointly on both tasks. In Table \ref{tab:cross} (Lower) we test this hypothesis by doing model selection on the average prevision between the two tasks. However, this strategy performs poorly in \textit{both} tasks for BeerAdvocate and is dominated by the CC task in Last.fm and MovieLens. This closely matches the CC rows. In other words, by selecting a network model jointly on both tasks, we never select the suitable model for link prediction in \textit{any} of the three datasets.

\section{Empirical Analysis: Measuring Efficiency}
In this evaluation, we evaluate our model selection with respect to our proposed efficiency measure. We look closely at the properties of this measure and similar stability in model ranking stable model ranking. We finally demonstrate model significance testing and sensitivity to noise. 

\subsection{Bootstrap and Rank Stability}

\begin{figure}
\centering
  \includegraphics[width=\subfigwidth\columnwidth]{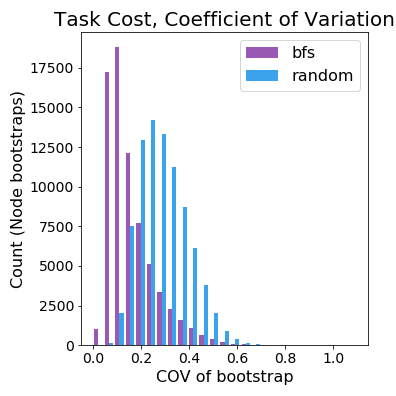}
   \includegraphics[width=\subfigwidth\columnwidth]{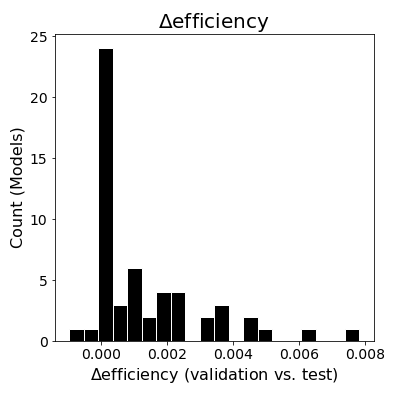}
  \caption{(Left) A distribution of the coefficient of variation ($\mu/\sigma$) over $b=20$ samples for a particular $\classifierset$ set of classification instances, for \lfm{} \social{}. This estimates the median of $b=100$ with low error ($\leq 0.02$). (Right) The change in efficiency between validation and testing partitions: $\mathcal{E}(\networkquery_{r, test})-\mathcal{E}(\networkquery_{r,validation})$.} 
  \label{fig:bootstrap}
\end{figure}

We now test the stability of our choice of node weighting, over $b=20$ re-sampled predictors. These replicates estimate the median of $b>100$ distribution with low error ($\leq 0.02$), therefore we proceed with $b=20$ for all results. Figure \ref{fig:bootstrap} (Left) reports the coefficient of variation ($\mu/\sigma$) of the encoding costs for the predictors in $\classifierset$. These are smaller for the \social{}-\bfs{} model on \lfm{} than for \social-\random. Figure \ref{fig:bootstrap} (Right) reports the signed difference in efficiency between models on the test and validation partition. The models with $\Delta$efficiency $> 0.004$ (or an increase of 250 bytes/correct) are all \bfs{} on \ml{}, which may be a legitimate change in efficiency, i.e. this is a poor performing set of models in test. This shows that our measurements are robust for estimating encoding cost at $k$, and efficiency is largely stable across partitions.

We test the stability of $\efficiency$ and its correlation to correct predictions. Table \ref{tab:kendall_e_vs_correct} (Top) reports the Kendall's $\tau$ rank order statistic measuring correlation between the `efficiency' ranking of models. This further shows stability in the models between validation and testing partitions. Table \ref{tab:kendall_e_vs_correct} (Bottom) in contrast reports the $\tau$ between the ranked models according to efficiency vs. correct predictions. Within the same partition. This rank correlation across measures is lower than the efficiency rank correlation \textit{across} partitions. This means that efficiency is quite stable in absolute error and relative ranking, and efficiency ranking is not merely a surrogate for ranking by correct predictions. Otherwise, we would not need to encode the model cost at all. 

\begin{table}
\centering
\resizebox{\boxscale\columnwidth}{!}{
\begin{tabular}{|c|c|c|c|}
\hline
\multicolumn{4}{|c|}{$\tau$, $\efficiency$, Validation vs. Test} \\
\hline
&\lfm{} & \ml{} &\ba{} \\
\hline\hline
$\tau$ & 0.89 & 0.61 & 0.82 \\
$p$-value & 1e-8  & 5e-4 & 3e-6 \\
\hline\hline
\multicolumn{4}{|c|}{$\tau$, $\efficiency$ vs. $\taskcorrect$ (validation)} \\
\hline
$\tau$ & 0.73 & 0.38 & 0.70 \\
$p$-value &3e-6  & 0.03 & 7e-5 \\
\hline
\end{tabular}}

\caption{(Top) The Kendall's $\tau$ rank correlation coefficient between model efficiency of validation vs. test partitions. (Bottom) $\tau$ rank correlation between efficiency and correct predictions.}
\label{tab:kendall_e_vs_correct}
\end{table}
\subsection{Efficiency Features}
Our definition of efficiency yields interpretable features that can characterize models and compare datasets. First, we compare the network and predictor encoding cost by calculating the ratio of the network encoding cost to total cost. 

Second, recall that in the efficiency definition, we select the most efficient $\kappa_i$ per node $i$, with $\kappa_i \leq 150$ in our evaluation. This gives the model the flexibility to explore wider in the network for each $i$. Nodes with higher $\kappa_i$ are likely harder to classify because the predictor encoding is typically more expensive when training over more attribute-vector instances. 

\begin{figure}
\centering
  \includegraphics[width=\figwidth\columnwidth]{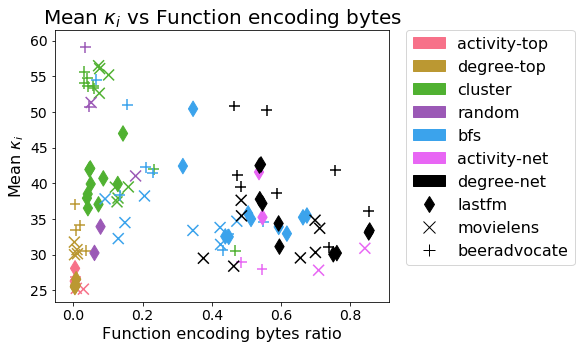}
  \caption{The mean over all $\kappa_i$: $\mathrm{mean}_i(\kappa_i)$ vs. the ratio of model cost by the total encoding cost: $\netcost/(\netcost+\taskcost)$. Colored by node weight function, with markers by dataset.} 
  \label{fig:scatter}
\end{figure}

Figure \ref{fig:scatter} compares the ratio of network and total encoding costs (x-axis) vs. the mean over all $\kappa_i$ (y-axis). The models roughly order left-to-right on the x-axis according to Table \ref{tab:functions}. 

All non-network models (\activityflat, \degreeflat, \cluster, \random) are consistently inexpensive vs. their predictor encoding cost, but several also select a higher $\kappa_i$. In contrast, the \degree{} and \activity{} models are consistently costlier because they encode a fixed $m > \max(k)$ in order to sample at each value $k$. All ordered weighting methods (network and non-network) tend to be most efficient at a lower $\kappa_i$. This may indicate these ordered nodes are more informative per instance, and each additional instance is more costly to incorporate into predictors, so our model selection will favor minimizing $\kappa$.

\subsection{Model Selection: Efficiency}

We now focus on selecting models from the efficiency ranking in validation. We've already shown that there is high stability in the efficiency ranking between validation and test, but less correlation between correct predictions and efficiency (Table \ref{tab:kendall_e_vs_correct}). Therefore, in this evaluation, we see whether we can recover the \textit{best} model in test ($\networkquery_{best}$) with respect to correct predictions, using efficiency selection criteria. 

\begin{table}[ht]
\centering
\resizebox{\boxscale\columnwidth}{!}{
\begin{tabular}{|l|c|c|c|}
\hline
\multicolumn{4}{|c|}{Model Selection: $\efficiency$, Evaluation: $\correct(C_r)$} \\
\hline
\multicolumn{1}{|c|}{$\networkquery_{select}$ (validation)}&$\frac{\mathcal{E}(\networkquery_{select})}{\mathcal{E}(\networkquery_{best})}$& $\frac{\cost(\networkquery_{select})}{\cost(\networkquery_{best})}$ & $\frac{\correct(\networkquery_{select})}{\correct(\networkquery_{best})}$ \\
\hline
\hline
\multicolumn{4}{|c|}{\lfm{}, $\networkquery_{best}$: \social-\bfs} \\
\hline

1. \social-\bfs & 1.00 & 1.00 & 1.00 \\
2. \social-\cluster & 0.78 & 0.29 & 0.56 \\
3. \knns-\cluster & 0.72 & 0.08 & 0.50 \\
\hline

\multicolumn{4}{|c|}{\ml{}, $\networkquery_{best}$: \activityflat} \\
\hline

1. \activityflat & 1.00 & 1.00 & 1.00 \\
2. \knns-\bfs &0.12 & 15.79 & 0.28 \\
3. \activity & 0.20 & 130.14 & 0.91 \\
\hline

\multicolumn{4}{|c|}{\ba{}, $\networkquery_{best}$: \activity} \\
\hline

1. \activityflat & 1.20 & 0.01 & 0.86 \\
2. \activity & 1.00 & 1.00 & 1.00 \\
3. \knns-\cluster & 0.22 & 0.09 & 0.20 \\
\hline

\end{tabular}}

\caption{Model selection on `efficiency' ranking in validation ($\networkquery_{select}$, Column 1) compared to the best model in test ranked by correct predictions ($\networkquery_{best}$). We measure efficiency, encoding cost, and correct prediction ratios between these two models (Columns 2,3,4).}
\label{tab:kendall_e}
\end{table}

The left-most column of Table \ref{tab:kendall_e} shows the three top-ranked models ($\networkquery_{select}$) by efficiency in validation. The ratios report the relative performance of the selected model (evaluated in test) for efficiency, total encoding size, and correct predictions. The second column--efficiency ratio--is not monotonically decreasing, because model ranking by efficiency may be different in test than validation (Table \ref{tab:kendall_e_vs_correct}).

The selection by efficiency shows an intuitive trade-off between encoding cost and predictive performance. In \lfm{}, the \social{}-\bfs{} model performs best, but the inexpensive \social{}-\cluster{} model is an alternative. This is consistent with previous results which show \social{}-\bfs{} is much more predictive than network adjacency, and of other underlying network models \cite{mlg2017_13}. 

According to performance ratios, \activity{} is extremely preferred in \ml{} and \ba{}. This measures the extent that the most active users are also most informative in terms of efficiency in these two domains. This is a striking result. Simply, within both domains, one only needs to do a weighted sampling of users by activity, which performs better than group-wise or pairwise models.

\ml{} correctly selects \activityflat{}, and \activity{} is an order of magnitude costlier while maintaining similar correct predictions. On \ba{}, \activity{} performs best on correct predictions, but \activityflat{} is the better model since it preserves $0.86$ of correct predictions but is $0.01$ the encoding cost. This demonstrates that efficiency is preferred when the application favors a parsimonious model.


\begin{figure}
\centering
\includegraphics[width=\threesubfigwidth\columnwidth]{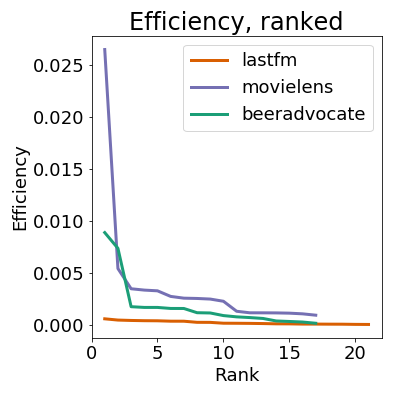}
\includegraphics[width=\threesubfigwidth\columnwidth]{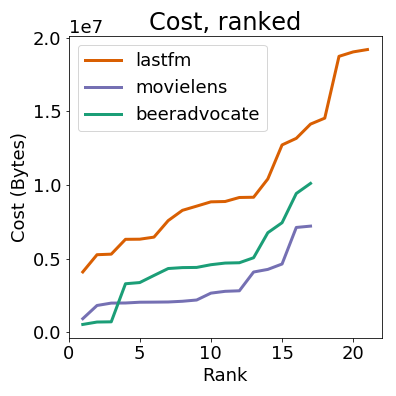}
\includegraphics[width=\threesubfigwidth\columnwidth]{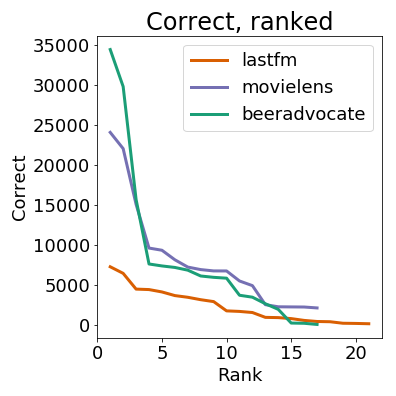}
\caption{Models ranked (x-axis) by efficiency (Left), total encoding cost $\taskcost + \netcost$ (Middle), and the number of correct predictions (Right).}
\label{fig:cost-correct}
\end{figure}

Figure \ref{fig:cost-correct} reports models ranked by efficiency (Left) total encoding cost (Middle), and correct predictions (Right). The \lfm{} ranking has more models because \social{} is included. Figure \ref{fig:cost-correct} (Middle) shows that each dataset grows similarly over our set of models, but that each has a different baseline of encoding cost. \lfm{} is particularly costly; the median non-zero attributes per node (e.g. artists listened) is 578, or 7 times larger than \ml{}. These larger attribute vectors yield more expensive predictor encoding, requiring more bytes for the same correct predictions. These baselines also yield the same dataset ordering in efficiency. So, the \textit{worst} model on \ml{} (\random{}, 1084 bytes/correct) is more efficient than the \textit{best} model on \lfm{} (\social{}-\bfs{}, 1132 bytes/correct). Figure \ref{fig:cost-correct}(Right) shows the extent that the first two ranked models (\activity{} and \activityflat{}) dominate the correct predictions on \ba{} and \ml{}. 


\subsection{Model Significance}
In order to compare all models for a dataset, we measure the \textit{significance} of each model relative to the efficiency over the complete set of models. 

Let $\networkquery_r$ be a model we intend to measure. We compare the median difference of efficiency of $\networkquery_r$ to all other models against the median pairwise difference of all models excluding $\networkquery_r$, normalized by the inter-quartile range (IQR) of pairwise difference. This measure is a non-parametric analog to the $z$-score, where the median differences deviate from the pairwise expectation by at least a `$\lambda$' factor of IQR. 

Let $e_i = \mathcal{E}(\networkquery_i)$, the efficiency value for an arbitrary model  $\networkquery_i$, and $\lambda$ a significance level threshold, then for $i = 1...|F|, j = 1...|F|; i, j \neq r$:\small{}
\begin{multline}
\mathrm{significance}(F, r, \lambda) = \\ \frac{\median_i(|e_r - e_i|) - \median_{i,j}(|e_i- e_j|)}{\iqr_{i,j}(|e_i - e_j|)} \geq \lambda 
\label{eq:zscore}
\end{multline}
\normalsize{}This is a \textit{signed} test favoring a larger efficiency of $\networkquery_r$ than the expectation. The median estimates of the $e_r$ comparisons and pairwise comparisons are also robust to a small number of other significant models, and like the $z$-score, this test scales with the dispersion of the pairwise differences.

This test only assumes `significant' models are an outlier class in all evaluated models $S$. This is a reasonable assumption because if many models efficiently represent the data, it may be a trivial application: use any network definition.  

At $\lambda=1$ in validation, we find five significant models, corresponding to models reported in Table \ref{tab:kendall_e}: \social{}-\bfs{} on \lfm{} ($=1.30$),  \activityflat{} and \activity{} on \ml{} ($=12.29, 1.21$) and \activityflat{} and \activity{} on \ba{} ($=7.64,~6.00$). 

\subsection{Model Robustness to Noise}

For each of the significant models, we measure the impact of noise on its efficiency and significance. We apply node rewiring on the model we are testing, and leave all other models intact. For each neighbor of node $i$, we rewire $i$ to a random node at probability $p$. This is out-degree-preserving randomization, which is appropriate because outgoing edges determine the input for the predictor on $i$. Sorting heuristics \activity{} and \degree{} are implemented as ad-hoc networks, where each node $i$ has directed edges to some $m$-sized subset of top-ranked nodes. Rewiring is done in the same manner on these networks, again to any node in $V$.

\begin{figure}
\centering
\includegraphics[width=\subfigwidth\columnwidth]{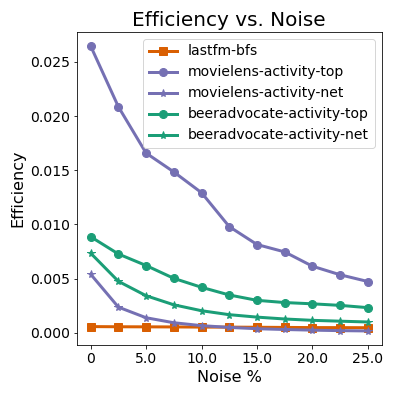}
\includegraphics[width=\subfigwidth\columnwidth]{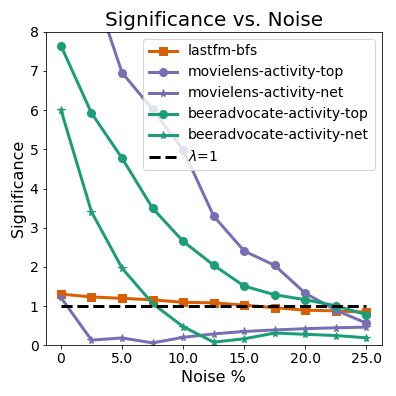}
\caption{Varying level of noise for 5 `significant' models (x-axis), reporting Efficiency (Left) and Significance (Right)}
\label{fig:significance_noise}
\end{figure}

Figure \ref{fig:significance_noise} shows the effect of varying noise $p$ (x-axis), on the efficiency (Left), and significance (Right) of each significant model. \activity{} on both \ml{} and \ba{} quickly lose efficiency under even small noise, and are no longer significant for $\lambda=1$ at $p=0.025$, and $p=0.10$, respectively. \activityflat{} is more robust, remaining significant to $p=0.225$. 

\activity{} is particularly sensitive to noise due to decreased performance in \textit{both} encoding cost and correct predictions. At only $p=0.025$, correct predictions on \activity{} on \ba{} reduce by $15\%$ and encoding cost \textit{increases} by $31\%$. The encoding cost greatly increases because the cardinality of the set of unique nodes in the \activity{} representation is small with many repetitions in the sequence of destination nodes. When random nodes are added in the representation, they are near-unique node IDs, greatly increasing the set cardinality and reducing the compression ratio. \activityflat{} shows a similar increase, proportional to nodes rather than edges.

Finally, \social{}-\bfs{} is easily the most robust to noise. From a lower baseline, it remains significant to $p=0.15$. It loses only $35\%$ of its significance value at any noise level, while all other methods lose $>90\%$. This demonstrates that network models have robustness which might be desirable for further criteria in model selection. For example, our full methodology can be used to select for efficiency with significance at some noise level $p$.

\section{Conclusions and Future Work}

In this work, we have formulated a methodology for network model selection from raw data with a focus on task performance to represent the behavior of the underlying system. Using this methodology, we demonstrate two key points. First, different networks represent different tasks, and often no one network represents multiple tasks. Second, we often find that a task is more \textit{efficiently} represented by a non-network structure such as groups or exemplars. Both of these takeaways reinforce a larger point: often there is not ``the network'' representation, but various possible representations of varying utility.

Recent work in graph neural networks (GNNs)\cite{wu2019comprehensive} makes these challenges of network definition even more difficult. In this case, the downstream model is highly parameterized and costly to train. Biases in different representations may be prohibitively costly to measure over this added complexity. There is a great need for efficient end-to-end GNN models which learn implicit network structure from raw data. Furthermore, training such models over alternative, simpler representations from raw data may encourage parameter-efficient models which haven't been a particular focus of this area.

\ifCLASSOPTIONcaptionsoff
  \newpage
\fi



%

\renewcommand*{\UrlFont}{\rmfamily}


\bibliography{ieeebib}
\bibliographystyle{IEEEtranN}




%
\vskip -2\baselineskip plus -1fil
\begin{IEEEbiography}[{\includegraphics[width=1in,height=1.25in,clip,keepaspectratio]{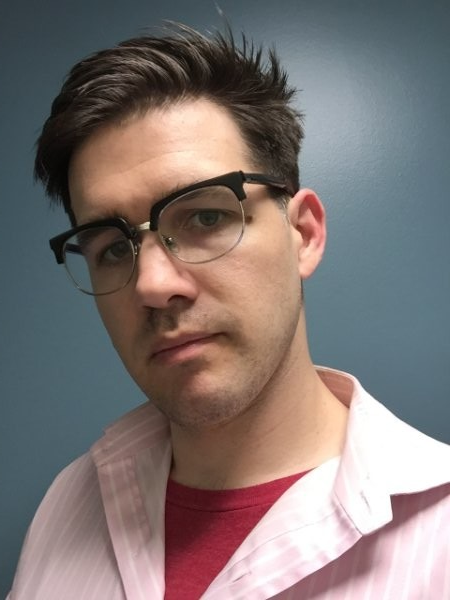}}]{Ivan Brugere}
Ivan Brugere is a PhD Candidate in the Department of Computer Science at the University of Illinois at Chicago and an NSF-IGERT Fellow in Electronic Security and Privacy. He holds an M.S. and B.S. from the University of Minnesota in Computer Science. 

His work focuses on statistical methodologies for inferring network structure from data in the application areas of computational ecology, neuroscience, and online social networks. Over his PhD, he held internships at Technicolor Research, Lawrence Livermore National Laboratory, Microsoft, and Amazon AWS. During that time, he received support from the Google Lime scholarship, NSF-IGERT program in Electronic Security and Privacy, and the University of Illinois at Chicago Chancellor's Graduate Research Fellowship.
\end{IEEEbiography}
\vskip -2\baselineskip plus -1fil
\begin{IEEEbiography}[{\includegraphics[width=1in,height=1.25in,clip,keepaspectratio]{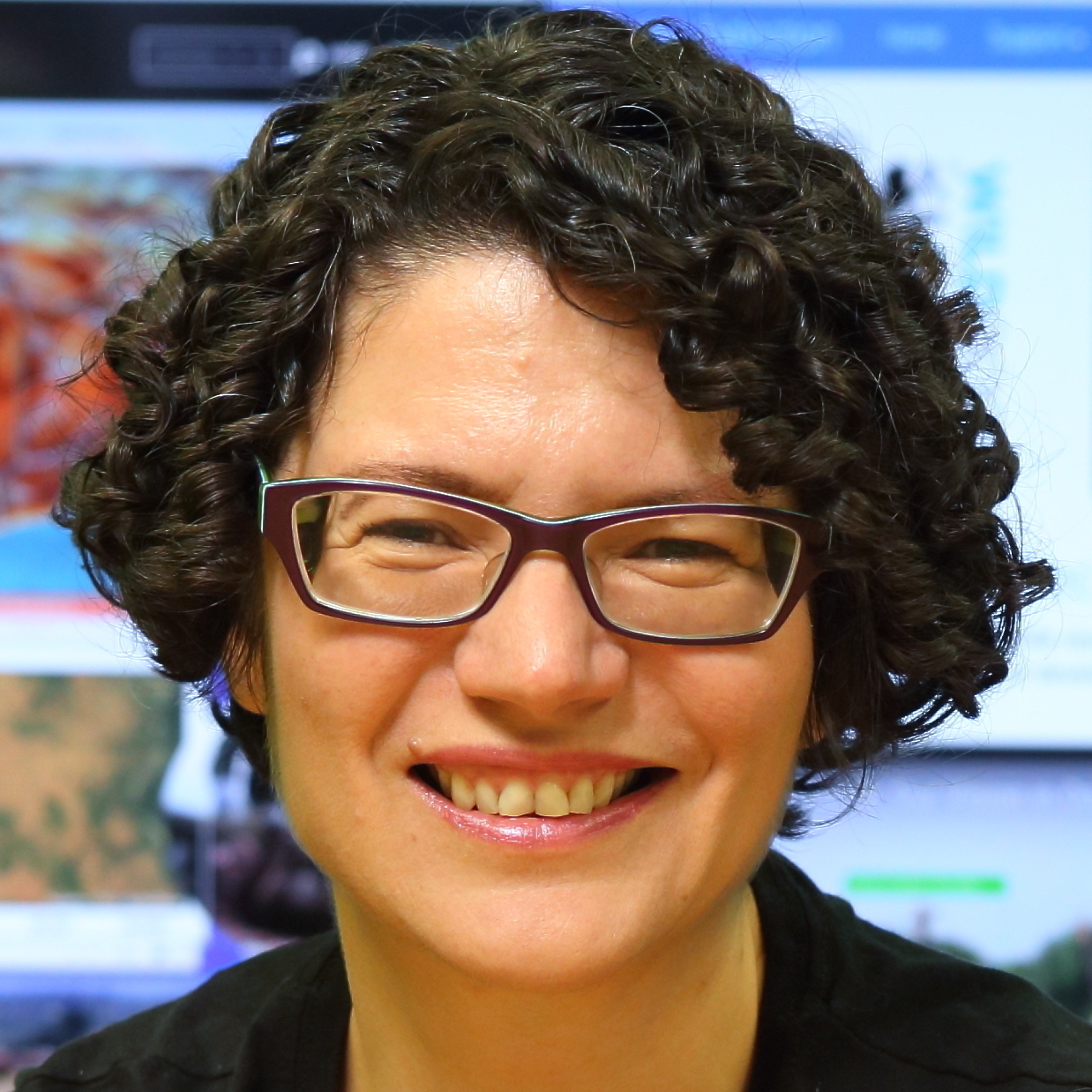}}]{Tanya Berger-Wolf}
Dr. Tanya Berger-Wolf is a Professor of Computer Science Engineering, Electrical and Computer Engineering, and Evolution, Ecology, and Organismal Biology at the Ohio State University, where she is also the Director of the Translational Data Analytics Institute.  Prior to The Ohio State University, she was at the University of Illinois at Chicago, where she directed the Computational Population Biology Lab in the Department of Computer Science. 

As a computational ecologist, her research is at the unique intersection of computer science, wildlife biology, and social sciences. She creates computational solutions to address questions such as how environmental factors affect the behavior of social animals (humans included). Berger-Wolf is also a director and co-founder of the AI for wildlife conservation non-profit Wild Me, home of the Wildbook project. 

Berger-Wolf holds a Ph.D. in Computer Science from the University of Illinois at Urbana-Champaign. She is widely published and is a sought-after invited speaker. She has received numerous awards for her research and mentoring, including University of Illinois Scholar, UIC Distinguished Researcher of the Year, US National Science Foundation CAREER, Association for Women in Science Chicago Innovator, and the UIC Mentor of the Year.
\end{IEEEbiography}




\end{document}